# Preparation of hexagonal iron flakes with a hexagonal structure on the sublayer of copper oxides

S. S. Parhizgar[1], Z. Ardeshiri

Nano lab, Plasma physics center, science and research branch Islamic Azad university, Tehran, Iran

**Abstract**. This study presented a novel method to prepare hexagonal iron flakes with hexagonal close-packed structures (hcp) on the copper oxide sublayers. Also, it demonstrated that the size and structure of iron grains are functions of sublayer features. The sublayer was prepared by sputtering copper on the glass substrate and annealing it. As the annealing temperature of the copper sublayer increased, the oxygen percentage increased and copper oxide rods were formed. Hexagonal iron flakes were formed after sputtering iron on this sublayer. The XRD analysis showed that these hexagonal flakes have a close-packed hexagonal structure. This finding was further confirmed by the magnetic properties of the samples and by density functional theory (DFT) calculations.

## I. INTRODUCTION

The physical properties of iron are considered to be of great importance in many fields in the sciences, as iron is one of the most abundant elements in the universe and the very basis for science and technology. Iron is polymorphic and can exist in body-centered cubic (bcc), face-centered cubic (fcc), and hexagonal close-packed (hcp) structures [1]. At atmospheric pressure, the bcc structure of Iron is stable below 910°C, and above 910°C, the iron structure changes to fcc structure. Here, an import question remains to be answered, i.e., can iron have a fcc structure at room temperature? The answer is yes. The fcc structure of iron at room temperature and atmospheric pressure is seen in the bilayer or multilayer of iron/x (x could be metal or semiconductors). As an example, thin iron films with a fcc structure, ~ 30 Å thick, have been grown on a (110) copper single crystal sublayer [2].

Bancroft et al. reported that the bcc phase of iron transits to the hcp structure at high pressure [3]. The reported transition pressure is in the range of 5 to 15 GPa [3,4]. The bcc-hcp phase transition has been studied using several theoretical [5,6] and experimental techniques [7-10] since it was first recognized. Another important question remains to be answered, i.e., can iron have the hcp structure at atmospheric pressure and room temperature? The answer is Yes. As J.P. Sanchez et. al. reported, Fe has the hcp structure in the Fe/Ru multilayer [11]. Also, the hexagonal flake form of Fe has been prepared, as reported by Li-Shun Fu et al. They obtained the hexagonal micro-flakes of Fe with a bcc structure by reducing $Fe_2O_3$ hexagonal micro-flakes [12].

In this article, first, the copper layer was sputtered on a glass substrate and annealed at different temperatures. Next, the iron layer was deposited on these copper oxide layers. The iron film prepared in this research exhibited the hexagonal flake form with hcp structure, which has not been done so far.

---

[1] contact author: s.parhizgar@srbiau.ac.ir

## II. METHODS
### 1. *copper and copper oxide sub-layer preparation*

In this paper, 500 nm thick copper thin films were deposited on the glass substrate using the DC magnetron sputtering method. A high purity (99.99%) Cu was used as the target. Before the deposition, the substrates were ultrasonically degreased with acetone for 10min, and then they were air dried. Initially, the chamber reached a base pressure of $10^{-4}$ Torr using rotary and turbomolecular pumps. 99.99% pure argon was entered into the chamber providing a working pressure of $10^{-2}$ Torr. During all deposition procedures, the chamber was kept at room temperature and after sputtering, the samples remained in the chamber without breaking the vacuum for 4h to facilitate the lattice relaxation and prevent the sample oxidation. This sample (Cu film) was named Cu. After preparing the Cu film, the annealing was carried out for thirty minutes at temperatures of 250, 350, and 450 °C in the pure argon atmosphere with $10^{-1}$ Torr pressure. The samples were labeled based on the annealing temperature. The samples annealed at 250, 350, and 450 °C were called Cu (250), Cu (350), and Cu (450) respectively.

### 2. *Iron layer preparation*

At this step, an iron layer was deposited on the sublayers, which were prepared in the previous step using a magnetic sputtering device. The deposition time on the different copper substrates was 40 minutes. The layer deposition conditions were repeated similar to the previous step. However, 99.99% purity of iron was used as the target this time. The samples were labeled based on the sublayer name, and the samples deposited on Cu(250), Cu(350), and Cu(450) were called Fe/Cu(250), Fe/Cu(350), and Fe/Cu(450) respectively.

Finally, Field Emission Scanning Electron Microscope (FESEM, TE-SCAN), X-ray diffraction technique (XRD, STADI MP, λ = 1.54 Å) and Vibration Sample Magnetometer (VSM, Lake Shore Cryotronics) were used to characterize the produced samples.

### 3. *DFT calculations*

The DFT calculations were performed in iron systems with hcp ($Fe_{hcp}$) and bcc ($Fe_{bcc}$) structures by using the projector augmented wave (PAW) method implemented in Quantum Espresso code. The 3d and 4s orbital electrons of Fe were adopted as valence electrons. A generalized gradient approximation (GGA) [18] was adopted for the exchange-correlation functional in the Perdew-Burke-Ernzerhof (PBE) parametrization. The lattice constant parameters were obtained from the XRD results. Therefore, the lattice constant values adopted for $Fe_{hcp}$ were a = 2.45 Å and b = 3.93Å, in p63/mmc space group, and for $Fe_{bcc}$ was a = 2.92 Å, in 229 IM-3M space group. The energy cutoff was set at 1000 eV for norm-conserving pseudopotential through the convergence test in our calculations, and the Monkhorst–Pack method was used to set the k-point mesh. The k-point separation was set at 0.03 (2 pi/a). The self-consistent field (SCF) tolerance was set at $5\times10^{-7}$ eV/atom. The energy calculation, for $Fe_{hcp}$ and $Fe_{bcc}$, at these sets for ferromagnetic and antiferromagnetic spin configurations was implemented to investigate the magnetic behavior. Next, the energy calculation was performed for the stable configuration to find the electronic properties.

## III. RESULTS & DISCUSSION
### 1. *Morphological and structural properties*

The samples were characterized by using the XRD technique. The X-ray diffraction patterns of Cu, Fe, and Fe/Cu samples are shown in Figure 1. In the XRD pattern of the Fe sample (Figure (a)), the lack of peak resulted from the amorphous structure of Fe. In Figure 1 (b), the XRD pattern of Cu shows a peak caused by planes [111] in the fcc structure with a lattice constant of 3.61Å (Ref. Code: 128-101-1016). After the Fe deposition on the Cu film, as shown in Figure 1(c), the diffraction peak of Fe/Cu increased compared to Cu, whereas if the Fe layer was in the amorphous structure, similar to the Fe sample, this peak should have decreased. This peak matches with peak (111) for iron in the fcc crystal structure with a lattice constant of 3.6Å and peak (110) in the bcc crystal structure with a lattice constant of 2.92Å. The lattice constant of 3.61Å for copper is close to the lattice constant of Fe with an fcc crystal structure. Several studies [13-16] demonstrated that the first few layers of Fe were formed in the fcc crystal structure because of the similarity of Cu and Fe lattice parameters and then continued to grow in the bcc crystal structure.

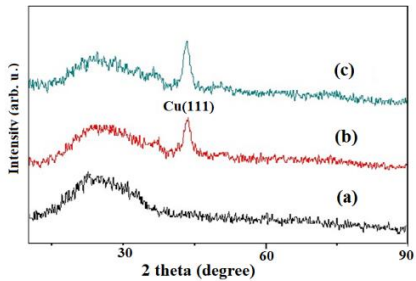

*FIG. 1 XRD patterns of samples (a) Fe (b) Cu and (c) Fe/Cu.*

Figure 2 shows the XRD patterns of samples Cu(250) and Fe/Cu(250); In this XRD pattern of Cu(250), the diffraction peaks (111) and (211) belong to the fcc structure of Cu, which is in agreement with the standard data reference (Ref. Code: 128-101-1016). Also, this pattern represents the diffraction peak related to the [-202] planes of CuO with a monoclinic structure (Ref. Code: 00-005-0661). In this pattern, the peak at $2\theta = 37.63°$ corresponds to $Cu_2O$ (Ref. Code: 00-035-1091). So, the annealing of Cu film leads to the oxidization of copper. In the XRD patterns of samples, Fe/Cu(250) is similar to the Cu(250) pattern, but the peak intensity decreased because of the amorphous iron layer deposited on Cu(250). The amorphous structure might be due to the crystallites with a bcc structure, which were too small that the X-ray analysis did not exhibit any sharp constructive peaks.

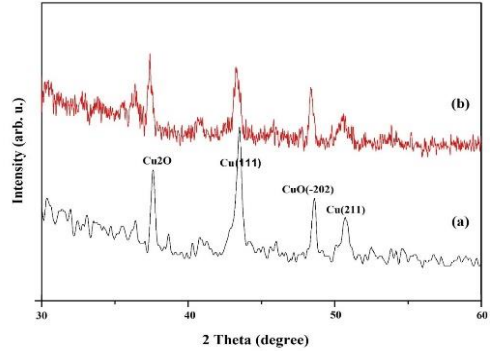

FIG. 2 XRD patterns of samples (a) Cu(250) and (b) Fe/Cu(250).

Figure 3 shows the XRD patterns of samples Cu(350) and Fe/Cu(350). According to the XRD pattern of sample Cu(350), Figure 3(a) shows [112], [-112] and [-202] plane diffraction peaks of CuO with a monoclinic structure (Ref. Code: 00-005-0661). Also the peak (111) of Cu with a fcc structure is seen in this pattern, which is decreased in comparison to the sample Cu(250) because the higher temperature led to the more oxidization of Cu. In the XRD pattern of the sample Fe/Cu(350), the diffraction peak (002) at $2\Theta = 46.03°$ (Ref. Code: 00-034-0529) is related to iron with a hcp structure in the p63/mmc space group with the lattice constant parameter of a = 2.45Å and b = 3.93Å. *The hexagonal structure of iron at atmospheric pressure is considered to be the key point within this study.*

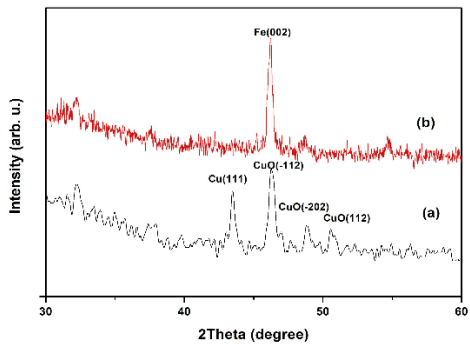

FIG. 3 XRD patterns of samples (a) Cu(350) and (b) Fe/Cu(350).

Figure 4 shows the XRD patterns of samples Cu(450) and Fe/Cu(450). According to the XRD pattern of the sample Cu(450), diffraction peaks (-112) and (110) are related to CuO. The peaks belonging to Cu are not seen, which indicate the complete oxidation of the sample, recognized by the XRD penetration depth. The XRD pattern of sample Fe/Cu(450), similar to Fe/Cu(350), shows the hcp structure of iron with the preferred texture of [002] planes.

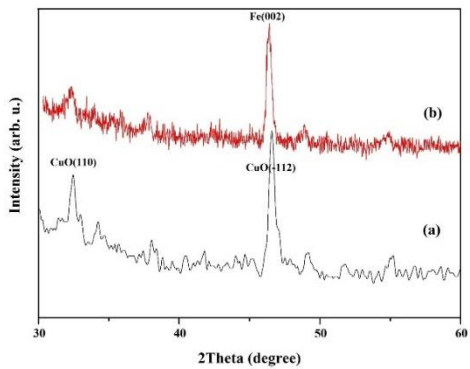

FIG. 4 XRD patterns of samples (a) Cu(450) and (b) Fe/Cu(450).

The FESEM images of samples Fe/Cu(250), Fe/Cu(350), and Fe/Cu(450) are shown in Figures 5-8. The thin layer of Au was sputtered on the surface samples before working with FESEM to obtain a good-quality FESEM image.

As presented in Figure 5, Fe/Cu(250) has grains with a spherical shape with an average diameter of 350nm, as seen in the image plane, while Cu(250) has such small grains that a uniform surface is seen in this scale.

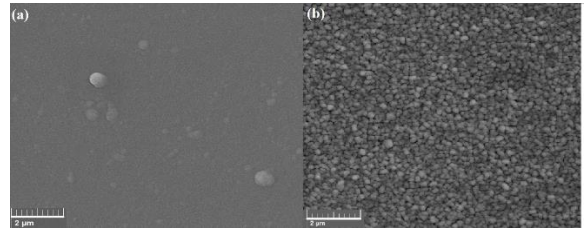

FIG. 5 FESEM images of samples (a) Cu (250) and (b) Fe/Cu(250).

Figure 6(a) and (b) show the FESEM images of samples Cu(350) and Fe/Cu(350), respectively. By annealing the copper layer and increasing it to 350ºC, in the sample Cu(350). the white-colored rods appeared (in Fig. 6a) related to copper oxide rods, which are in agreement with other reports [17-20]. Figure 6b shows that stuck hexagonal iron sheets with an average thickness of 231 nm have grown on the copper oxide sublayer. The formation of this shape of iron sheets can be due to the oxidation of copper surface and due to the reduction of surface energy compared to copper. A lower surface energy reduces the adhesion of iron onto the sublayer. Also, in sputtering on the columnar structure sublayer, the tips of the columns are the site of the nucleation seeds. The growth of the seeds continues as the sputtering continues. Then the smaller the number of grains, the larger the islands are formed.

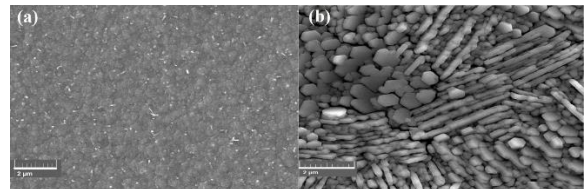

FIG. 6 FESEM images of samples (a) Cu (350) and (b) Fe/Cu(350).

Figure 7 (a and b) shows the FESEM images of samples Cu(450) and Fe/Cu(450), respectively. Because of increasing annealing temperature in the sample Cu(450), the grains grew compared to Cu(350). Also, the wider valley between the grains shows the vertical growth of grains. In this figure, the iron grains have grown in discrete flakes at 450ºC, while at 350ºC, they are not completely discrete and make large sheets. As shown in Figure 7b, the hexagonal flakes with an average diameter of 435 nm are formed. In addition to hexagonal flakes, triangular flakes are also seen in the Fe/Cu(450) sample. The more uniform flake sizes in the Fe/Cu(450) sample were observed compared to the Fe/Cu(350) sample because of increasing the annealing temperature of the sublayer (Cu), thus leading to more copper oxide rods. Then the higher number and the more uniform nucleation sites in sample Cu(450) led to higher and the more uniform iron flakes of Fe, which are smaller in size compared to Fe/Cu(350).

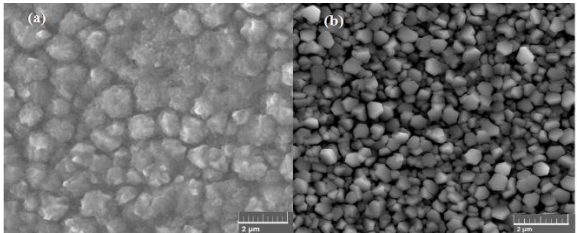

FIG. 7 FESEM images of samples (a) Cu(450) and (b) Fe/Cu(450)

### 2. Electronic and magnetic Properties

The DFT energy calculations for $Fe_{hcp}$ and $Fe_{bcc}$ for ferromagnetic and antiferromagnetic spin configurations were performed to estimate the magnetic behavior. These calculations show that Fe in the hcp structure with an antiferromagnetic configuration has less free energy than the ferromagnetic configuration, and Fe in the bcc structure with a ferromagnetic configuration has less free energy than the antiferromagnetic configuration. Therefore, $Fe_{hcp}$ prefers to be antiferromagnetic and $Fe_{bcc}$ prefers to be ferromagnetic.

The electron density of states of $Fe_{hcp}$ and $Fe_{bcc}$ have been calculated to understand their magnetic and electronic properties. Figure 8(a & b) illustrates the bulk electronic DOS with spin-up (black) and spin-down (red) of iron in bcc ($Fe_{bcc}$) and hcp ($Fe_{hcp}$) structures. The Fermi energy were set to zero and the DOS values were calculated in the interval [–20, 20] eV. As shown in this figure, the density of valence electrons of $Fe_{bcc}$ with spins down and spin up is different; the difference between areas of spin up and spin down is the bulk magnetic moment. The magnetic moment obtained for $Fe_{hcp}$ is almost zero and it is $2.64 \mu_B\, per\, atom$ for $Fe_{bcc}$, which is in agreement with other reports [21].

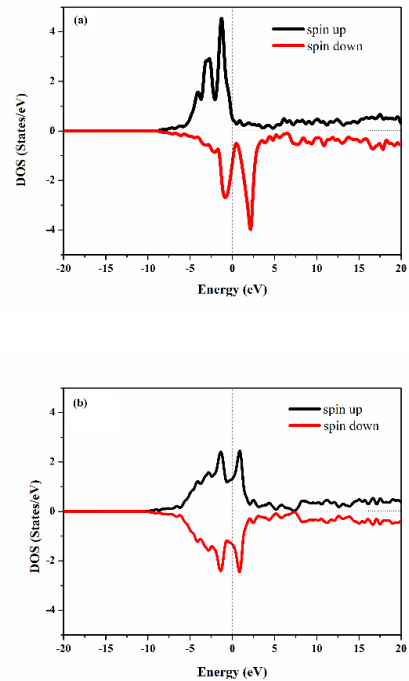

FIG. 8 DFT $DOS_S$ of iron bulk in (a) bcc structure and (b) hcp structure

Figure 9 illustrates the normalized magnetization (M/M$_S$) versus in-plane applied magnetic field (H) of the samples Fe/Cu(250) and Fe/Cu(450). M is the sample magnetization as a function of H and $M_S$ is the saturation magnetization of sample Fe/Cu(250). This figure shows the effect of sublayer features on the magnetization of the iron layer. It should be noted that the magnetization of the sublayer is negligible for two samples because the XRD analysis shows that the Cu(250) substrate is composed of Cu, CuO, and $Cu_2O$, and the Cu(450) substrate is composed of CuO. The bulk pure CuO and $Cu_2O$ show a paramagnetic nature, while their nanowires show a ferromagnetic nature [22,23]. Besides, the bulk CuO/$Cu_2O$ composites demonstrate a room temperature ferromagnetic nature [22]. But the saturated magnetization of CuO/$Cu_2O$ composites and nanowires of copper oxide composites are very small (in the range of [0.04-.06] emu/g) [22,23] compared to $Fe_{bcc}$ (55 emu/g). As illustrated in this figure, Fe/Cu(250) has more magnetization than Fe/Cu(450) at any applied magnetic field. The saturation magnetization of Fe/Cu(250) is ten times larger than Fe/Cu(450) magnetization at one Tesla.

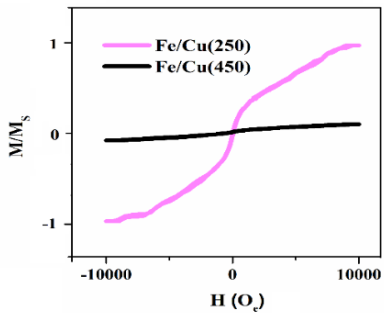

FIG. 9 Normalized magnetization M/MS vs applied magnetic field for Fe/Cu(250) and Fe/Cu(450)

The main reason for the less magnetization of the sampleFe/Cu(450) than the magnetization of the sampe Fe/Cu(250) is due to their different crystal structure. As the XRD results show, Fe/Cu(450) has a hcp structure. In the hcp structure, the distance between the nearest neighboring atoms is 2.421Å, and in the bcc structure, the distance between the nearest neighboring atoms is 2.529 Å. According to the Bethe–Slater curve, the sign of the exchange interaction changes when the distance between the atoms changes. It can be said that the reduction of the distance between the atoms has led to the antiferromagnetic behavior in this sample. This is consistent with the DFT calculations concluded in this study.

**IV. Conclusions**

In this study, hexagonal iron fakes were formed on the copper oxide sublayers. In this configuration, the copper sublayer was first made by DC magnetron sputtering, and then the samples were annealed at different temperatures to form the layers of copper oxides with different morphologies. Increasing the heating temperature caused an increase in the height size of the grains of the sublayer. After sputtering iron on the sublayer of copper oxides, hexagonal iron flakes were formed with the hexagonal structure on the copper oxide sublayers. These iron flakes with the hexagonal structure exhibited antiferromagnetic behavior, as confirmed by DFT calculations.